\documentclass[%
aps,
prx,
 amsmath,amssymb,
 superscriptaddress,
 preprint,%
reprint,%
]{revtex4-2}

\usepackage{graphicx}
\usepackage{dcolumn}
\usepackage{bm}
\usepackage{ulem}
\usepackage[utf8]{inputenc}
\usepackage[T1]{fontenc}
\usepackage{mathptmx}
\usepackage{textcomp} 
\usepackage{nicefrac}
\usepackage{xcolor}
\usepackage{gensymb}
\usepackage{braket}

\newcommand{\exx}{\ensuremath{\epsilon_{xx}}}
\newcommand{\exy}{\ensuremath{\epsilon_{xy}}}
\newcommand{\reexy}{\ensuremath{\mathrm{Re}(\epsilon_{xy})}}
\newcommand{\imexy}{\ensuremath{\mathrm{Im}(\epsilon_{xy})}}
\newcommand{\vexy}{\ensuremath{\vec{\epsilon}_{xy}}}
\newcommand{\ef}{\ensuremath{E_\textsubscript{F}}}
\newcommand{\ptheta}{\ensuremath{\vec{p}_\theta}}

\begin{document}

\title{Unraveling Femtosecond Spin and Charge Dynamics with EUV T-MOKE Spectroscopy
}

\author{Henrike Probst}%
\author{Christina Möller}%
\author{Maren Schumacher}
\affiliation{I. Physikalisches Institut, Georg-August-Universit\"at G\"ottingen, Friedrich-Hund-Platz 1, 37077 G\"ottingen, Germany}
\author{Thomas Brede}
\affiliation{Institute of Materials Physics, University of G\"ottingen, Friedrich-Hund-Platz 1, 37077 G\"ottingen, Germany}
\author{John Kay Dewhurst}%
\affiliation{Max-Born-Institute for Non-linear Optics and Short Pulse Spectroscopy, Max-Born Strasse 2A, 12489 Berlin, Germany}
\author{Marcel Reutzel} %
\author{Daniel Steil} %
\affiliation{I. Physikalisches Institut, Georg-August-Universit\"at G\"ottingen, Friedrich-Hund-Platz 1, 37077 G\"ottingen, Germany}

\author{Sangeeta Sharma} 
\affiliation{Max-Born-Institute for Non-linear Optics and Short Pulse Spectroscopy, Max-Born Strasse 2A, 12489 Berlin, Germany}

\author{G.~S.~Matthijs Jansen} \email{gsmjansen@uni-goettingen.de} %

\author{Stefan Mathias} \email{smathias@uni-goettingen.de}%
\affiliation{I. Physikalisches Institut, Georg-August-Universit\"at G\"ottingen, Friedrich-Hund-Platz 1, 37077 G\"ottingen, Germany}

\begin{abstract}
The 
magneto-optical Kerr effect (MOKE) in the extreme ultraviolet (EUV) regime has helped to elucidate some of the key processes that lead to the manipulation of magnetism on ultrafast timescales. However, as we show in this paper, the recently introduced spectrally-resolved analysis of such data can lead to surprising experimental observations, which might cause misinterpretations. Therefore, an extended analysis of the EUV magneto-optics is necessary. Via experimental determination of the dielectric tensor, we find here that the non-equilibrium excitation in an ultrafast magnetization experiment can cause a rotation of the off-diagonal element of the dielectric tensor in the complex plane. In direct consequence, the commonly analyzed magneto-optic asymmetry may show time-dependent behaviour that is not directly connected to the magnetic properties of the sample. We showcase such critical observations for the case of ultrafast magnetization dynamics in Ni, and give guidelines for the future analysis of spectrally-resolved magneto-optical data and its comparison with theory.
\end{abstract}

\maketitle

\section{Introduction}

With the advent and success of laser-based femtosecond element-specific M-edge magneto-optical Kerr spectroscopy in 2009 \cite{la-o-vorakiat_ultrafast_2009}, a very successful series of experiments using this technique has helped to verify and elucidate a number of key findings in ultrafast magnetism \cite{mathias2012probing, pfau2012ultrafast, rudolf2012ultrafast, vodungbo2012laser, turgut2013controlling, willems_optical_2020, hofherr_ultrafast_2020, tengdin_direct_2020, hennecke2022ultrafast, Kfir:2017gs, Zayko.2020,Ryan.2023}. As more powerful laser systems have become available over the years, this experimental technique is becoming available in a growing number of laboratories and becoming an important workhorse to study ultrafast magnetization dynamics \cite{jana2017setup, hofherr2018, yao2020tabletop, moller_ultrafast_2021, johnsen2023beamline}. However, the introduction of a new experimental capability also requires an ongoing check of the validity of the collected data, and a crosscheck of the results achieved with complementary experimental techniques. Such validation has been done, for example, for the case of photo-induced ultrafast spin currents and the heavily discussed relative delay of the Fe and Ni demagnetization in a Permalloy sample \cite{mathias2012probing, gunther2014testing, yao_distinct_2020, jana2017setup, moller_ultrafast_2021, jana2022experimental}.

Currently, spectrally-resolved analysis of femtosecond transverse magneto-optical Kerr effect (T-MOKE) data in the extreme-ultraviolet (EUV) is more frequently used, for example for the verification of the so-called optical intersite spin-transfer (OISTR) effect \cite{dewhurst2018, siegrist_light-wave_2019, hofherr_ultrafast_2020, tengdin_direct_2020, willems_optical_2020}. However, it has already been recognized that the interpretation of such data 
asks for a more detailed analysis of the collected T-MOKE asymmetry data \cite{la-o-vorakiat_ultrafast_2012, jana_analysis_2020}. Similarly, the comparison of magneto-optical spectroscopy data with state-of-the-art time-dependent density functional theory (TDDFT) and similar theoretical methods requires careful analysis \cite{willems_magneto-optical_2019, dewhurst2020element}. 

\begin{figure}[thb!]
\centering
    \includegraphics[width=0.9\columnwidth]{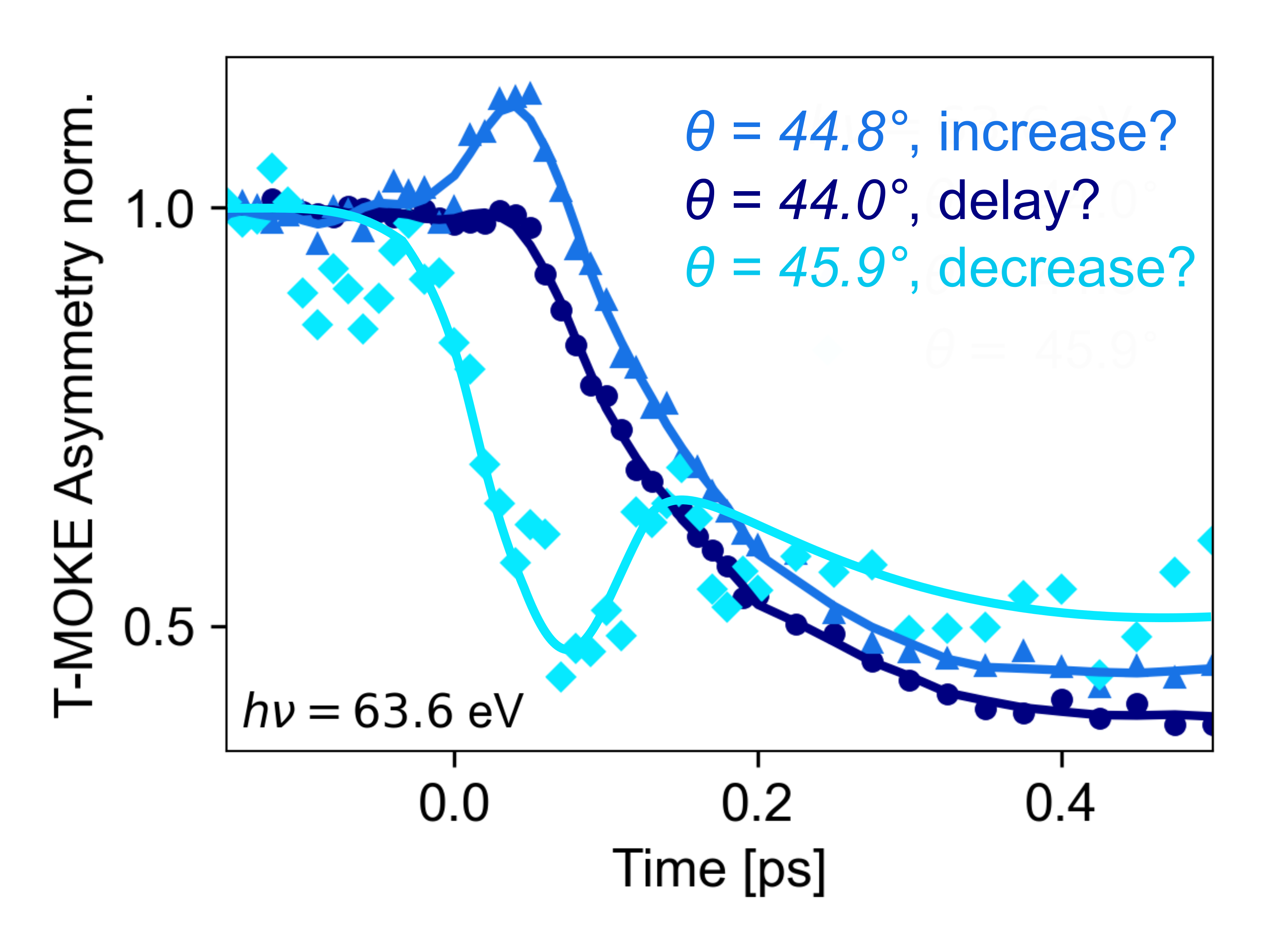}
    \caption{Increase, delay, or decrease of the spin dynamics? Very different dynamics in the T-MOKE asymmetry are observed at 63.6~eV ($\pm 2\%$) photon energy for slightly different angles-of-incidence (dark blue: 44\textdegree, blue: 44.8\textdegree, light blue: 45.9\textdegree). The measurement clearly illustrates that a direct interpretation of T-MOKE asymmetry with ultrafast magnetization dynamics can be highly problematic. Note that the solid lines serve as a guide to the eye.
    }
    \label{fig:dynamics_angles}
\end{figure}

The urgency for a sophisticated analysis of spectrally-resolved T-MOKE data can be best introduced with an exemplary measurement of ultrafast magnetization dynamics in the prototypical 3d ferromagnet Ni. Fig.~\ref{fig:dynamics_angles} shows transient T-MOKE asymmetry data, measured with our setup \cite{moller_ultrafast_2021}, and analyzed at 63.6~eV ($\pm 2\%$) EUV photon energy for slightly varying incidence angles. Very disturbingly, the transient dynamics of the T-MOKE asymmetry shows completely distinct behaviour. In particular, for the very same ultrafast pump-probe experiment, all heavily discussed experimental signatures, i.e. an increase, a delay, and a decrease in the T-MOKE asymmetry can be identified, which have previously been used to identify the OISTR effect \cite{hofherr_ultrafast_2020, tengdin_direct_2020, willems_optical_2020}, the influence of exchange scattering \cite{mathias2012probing}, and the typical demagnetization process \cite{la-o-vorakiat_ultrafast_2009}. Having just the information of this particular measurement available, all previous interpretations of T-MOKE data would be in question. In the following, we will show how such critical T-MOKE data needs to be analyzed to obtain reliable access to the true spin dynamics.

\begin{figure}[t]
\centering
\includegraphics[width=\columnwidth]{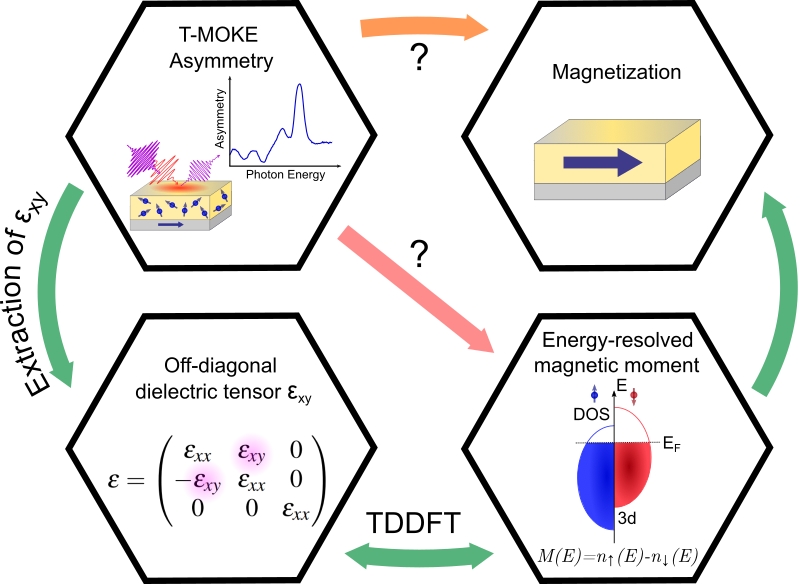}
\caption{Overview of the connections between experimental magneto-optical signal, the off-diagonal tensor-element and the magnetization.
}
\label{fig:scheme_Asym_to_Spinpol}
\end{figure}

The overall topic of the paper is illustrated in Fig.~\ref{fig:scheme_Asym_to_Spinpol}. Usually, the main quantity of interest is the global magnetization (top right) or the energy-resolved magnetic moment, given by the difference in majority and minority spins in the density of states (bottom right). In EUV T-MOKE experiments, the aim is to probe the time-dependent magnetization and magnetic moment, however neither magnetization (orange arrow) nor the energy-resolved magnetic moment (red arrow) are directly measured. Rather, as an optical technique, T-MOKE probes the dielectric tensor and is specifically sensitive to the off-axis dielectric tensor element \exy{} \cite{erskine_calculation_1975}. Since the dielectric tensor (including \exy{}) can be calculated from TDDFT calculations of the spin-resolved density of states \cite{runge_density-functional_1984, van_leeuwen_key_2001, sharma_optical_2014, dewhurst2020element, willems_magneto-optical_2019}, a time-resolved extraction of \exy{} from T-MOKE data would allow for a quantitative comparison of experimental T-MOKE data and theoretical calculations.

In our work, we therefore develop a robust and easy-to-implement method to analyze transient dynamics of the dielectric tensor. With the help of this analysis, we elucidate that the non-equilibrium excitation by the optical pulse can lead to a rotation of the off-diagonal element of the dielectric tensor in the complex plane. This rotational behaviour can lead to the observed increase, delay, and decrease of the T-MOKE asymmetry as seen in Fig.~\ref{fig:dynamics_angles}, and these differing signals simply depend on the used measurement geometry of the T-MOKE experiment, i.e. the angle of incidence of the EUV light.  As TDDFT is also able to provide transient dynamics of the dielectric tensor, we find that a direct comparison of experiment and theory becomes possible via such a dielectric tensor analysis from the data. In Ni, via comparison with theory, we can show that the observed dynamics at early times of the pump excitation (<50~fs) is dominantly driven by spin-conserving transitions in the minority channel. Besides a comparison of the same quantity, our approach also ensures that spectral broadening, multiple edges, and overlapping edges from multiple elements in multi-component materials is properly taken into account.

\section{Magneto-optical spectroscopy}
\label{sec:medge_spectroscopy}
\subsection{Magneto-optical asymmetry in T-MOKE}
\label{subsec:magnetic_asymmetry} 
Previous magneto-optical studies have attempted to make a direct connection between the magneto-optical signal and the element-specific magnetization properties of the sample after an ultrafast excitation. However, as shown above and discussed in Fig.~\ref{fig:scheme_Asym_to_Spinpol}, the relation between the measurement signal in a T-MOKE experiment and the magnetization is more complex, even in the equilibrium case. 
At the microscopic scale, the interaction of light with a material is given by the complex-valued dielectric tensor $\epsilon$, which can be derived by counting the number of allowed optical transitions within the the spin-resolved band structure \cite{oppeneer1992ab, oppeneer2001magneto, kunevs2001x, dewhurst2020element, krieger2015laser}. For a magnetic material, the imbalance between the different spin channels leads to off-diagonal terms in the dielectric tensor, which couple light fields of orthogonal polarization. For a typical (cubic) magnetic material that is magnetized along the $z$-axis, the dielectric tensor is commonly expressed as 
\begin{equation} \label{eq:dielectric_tensor}
\epsilon = 
    \begin{pmatrix}
    \exx{} & \exy{} & 0 \\
    -\exy{} & \exx{} & 0 \\
    0    & 0    & \exx
    \end{pmatrix}.
\end{equation}
Here, \exx{} can be directly related to the non-magnetic refractive index, mostly written as
\begin{equation}
\label{eq:refractive_index}
\sqrt{\exx} = n = 1-\delta + i \beta, 
\end{equation} 
while the off-axis dielectric tensor element \exy{} describes the magneto-optical response of the material.

Through the Fresnel equations, it is possible to express the signal in various magneto-optical techniques in terms of the dielectric tensor components. For example for XMCD, it is known that a signal can be extracted that is proportional to \reexy{} \cite{dewhurst2020element, willems_magneto-optical_2019}. For T-MOKE, the reflectivity of a single vacuum/magnetic material interface can to good accuracy be expressed as \cite{zusin2018direct} 
\begin{equation} \label{eq:intensity_up_down}
R_{\uparrow/\downarrow} = |R_0|^2 + |R_m \exy{}|^2 \pm \mathrm{Re}\{2R_0^*R_m\exy{}\}
\end{equation}
where $R_0 = \frac{n\cos{\theta_i} - \cos{\theta_t}}{n\cos{\theta_i} + \cos{\theta_t}}$ and $R_m = \frac{sin{2\theta_i}}{n^2(n\cos{\theta_i} + \cos{\theta_t})}$ are the (complex-valued) non-magnetic and magnetic contributions to the reflectivity, respectively. $\mathrm{Re}\{...\}$ indicates the real part, and the incidence angle $\theta_i$ and refraction angle $\theta_t$ are related by Snell's law. The $\pm$ sign in Eq.~\ref{eq:intensity_up_down} is directly linked to the magnetization direction, where switching the magnetization direction also switches the sign.

Based on Eq.~\eqref{eq:intensity_up_down}, the commonly-used measurement strategy in EUV T-MOKE is to measure the reflected intensity $I_{\uparrow/\downarrow}$ (which is proportional to $R_{\uparrow/\downarrow})$, and subsequently calculate the T-MOKE asymmetry for bulk magnetic material by \cite{dewhurst2020element, la-o-vorakiat_ultrafast_2012, jana_analysis_2020, mathias2013ultrafast}
\begin{equation} \label{eq:asym}
A =\frac{I_{\uparrow}-I_{\downarrow}}{I_{\uparrow}+I_{\downarrow}} = \frac{2\text{Re}(R_0^*R_m\exy)}{|R_0|^2+|R_m\exy|^2}.
\end{equation}
By alternately measuring $I_\uparrow$ and $I_\downarrow$, this allows to efficiently filter out variations in the light source intensity. Then, it is useful to consider the following assumptions: if $|R_m\exy{}|^2 \ll |R_0|^2$ and the refractive index $n$ is constant, then the transient asymmetry $A(t)$ can be normalized to a reference asymmetry $A_{\rm ref}$ (commonly the T-MOKE asymmetry of the sample in equilibrium), and the signal 
\begin{equation} 
    \frac{A(t)}{A_{\rm ref}} = \frac{\mathrm{Re}(R_0^*R_m\exy(t))}{\mathrm{Re}(R_0^*R_m\exy^{\rm ref})}
\end{equation}
is acquired. Now, a further assumption can be made to finally quantitatively link the observed signal $A(t)$ to a change in \exy{}: namely that the angle of \exy{} in the complex plane does not change and only the magnitude $|\exy{}|$ decreases/increases. However, as will become clear in the following analysis, the data that is presented in Fig.~\ref{fig:dynamics_angles} unambiguously shows that this set of assumptions cannot hold.

\subsection{Angle-dependence of the T-MOKE asymmetry}
\label{subsec:innerproduct}
It is well known that the reflectivity of a sample, and more specifically also the T-MOKE signal, depend strongly on the angle of incidence $\theta_i$ \cite{mathias2012probing, turgut_stoner_2016, hennecke2022ultrafast}. This is particularly true for the magneto-optical reflectivity close to the Brewster angle, 
where a strong magneto-optical signal is observed as the non-magnetic reflection is strongly suppressed. In order to understand this behavior, it is useful to consider $\exy{}=\reexy{}+i\,\imexy{}$ as a vector in the complex plane: \vexy{}. Approximating the T-MOKE asymmetry to depend linearly on \exy{} (i.e., using $|R_0|^2 \gg |R_m\exy|^2$ in Eq.~\eqref{eq:asym}), we rewrite Eq.~\eqref{eq:asym} as
\begin{equation} \label{eq:innerproduct_linear}
\begin{split} \nonumber
A(\exy{}) &\approx
\frac{2}{|R_0|^2} \cdot \text{Re}(R_0^*R_m\exy)\\
 & = \frac{2}{|R_0|^2} \cdot [\mathrm{Re}(R_0 R_m^*)\cdot\mathrm{Re}(\exy{}) + \mathrm{Im}(R_0 R_m^*)\cdot\mathrm{Im}(\exy{})] \\
 & = \vec{p}_{\theta} \cdot \vexy{} = |\vec{p}_{\theta}| |\vexy{}| \cos(\sphericalangle(\ptheta{}, \vexy{})).
\end{split}
\end{equation}
This analysis shows that it is possible to interpret the T-MOKE asymmetry $A(\vexy{})$ as inner product of \vexy{} with a \textit{probe vector} \ptheta{}. 
For the simple case of a single vacuum/magnet interface, \ptheta{} is (up to a scaling factor) proportional to $(\mathrm{Re}(R_0 R_m^*), \mathrm{Im}(R_0 R_m^*))$. More generally, we define the probe vector as the derivative of the T-MOKE asymmetry with respect to \exy{}:
\begin{equation} \label{eq:innerproduct}
    \ptheta{} = \left(\frac{\partial A}{\partial(\reexy{})}, \frac{\partial A}{\partial(\imexy{})}\right).
\end{equation}
We note that since \ptheta{} depends only on the geometric and non-magnetic properties of the sample, it can be calculated without (precise) a-priori knowledge of \exy{}. It is most important, however, to realize that \ptheta{} rotates strongly in the complex plane as the angle of incidence $\theta_i$ of the EUV light in the T-MOKE measurement changes.

\begin{figure}[htb]
\centering
    \includegraphics[width=1.0\columnwidth]{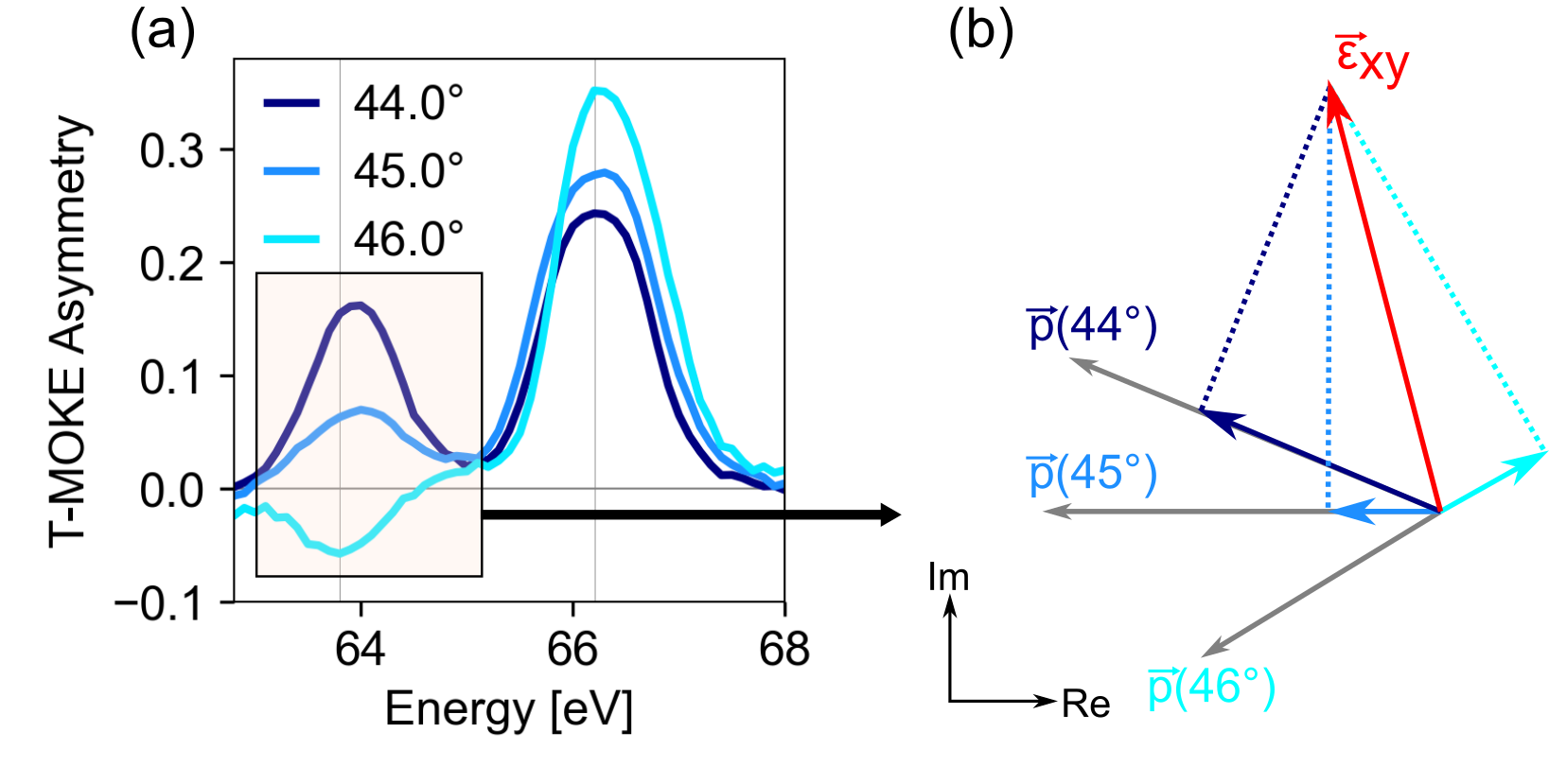}
    \caption{Angle-dependence of the static T-MOKE asymmetry for Ni. (a) The observed T-MOKE asymmetry for the 21~nm Ni sample for three different incidence angles $\theta_i$ close to 45\textdegree{}. (b) Schematic representation of the probe vectors following Eq.~\ref{eq:innerproduct} for the different incidence angles in (a), and the \exy{} that was determined from these measurements. 
    For the calculation, we used the same sample composition and refractive index values as in Section~\ref{sec:exy}. 
    }
    \label{fig:asymm_arrow}
\end{figure}

This strong $\theta_i$ angle of incidence dependence is critical, because the T-MOKE asymmetry is proportional to the inner product of the vectors \vexy{} and the probe vector \ptheta{}. If the angle $\sphericalangle(\ptheta{}, \vexy{})$ between these vectors approaches 90\textdegree{}, a particularly strong dependence of the asymmetry on the geometrical factors must be expected. This situation is illustrated in Fig.~\ref{fig:asymm_arrow}a, where the measured T-MOKE asymmetry for three incidence angles close to $\theta_i =$~45\textdegree{} is shown. In the spectral region marked with a rectangle, the T-MOKE asymmetry is strongly angle-of-incidence-dependent, and even flips from positive to negative values. This can be understood by looking at the schematic in Fig.~\ref{fig:asymm_arrow}b, which illustrates the positive to negative change of the inner product of $\ptheta{}$ and $\vexy{}$ as a function of the angle-of-incidence and therewith $\sphericalangle(\ptheta{}, \vexy{})$.

Another very important aspect for angles $\sphericalangle(\ptheta{}, \vexy{})\approx 90\degree$ is the realization that even small rotations of \exy{} in time-resolved measurements would lead to sign-changes in the measured T-MOKE asymmetry. Indeed, we will show that this is exactly the reason for the disturbing data presented in Fig.~\ref{fig:dynamics_angles}, which was analyzed at a photon energy of 63.6~eV: here, a small transient rotation of \exy{} leads to very different transient T-MOKE asymmetries for different angle-of-incidences, as we will fully analyze below.

\begin{figure}[thb]
\centering
    \includegraphics[width=0.9\columnwidth]{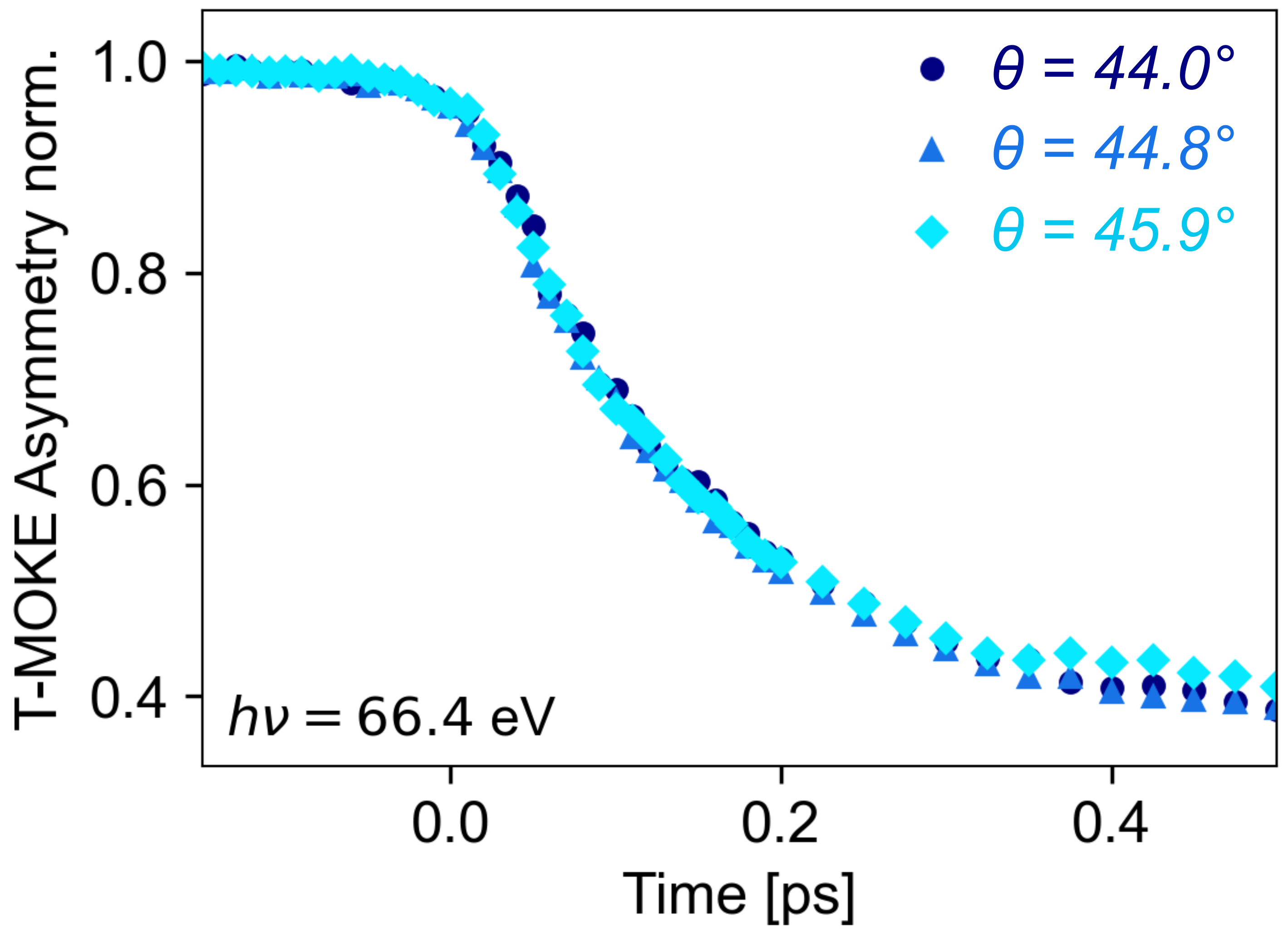}
    \caption{Transient magneto-optical asymmetry at 66.4~eV, i.e., at the Ni resonance, yields identical dynamics for each angle. }
    \label{fig:good_measurement}
\end{figure}

On the other hand, if $\sphericalangle(\ptheta{}, \vexy{})$ is much larger or smaller than 90$\degree$, as is the case in the spectral region around 66~eV, no such peculiar behaviour is expected. An analysis of transient T-MOKE asymmetry around 66~eV for different angles of incidence consequently yields perfectly identical results (Fig.~\ref{fig:good_measurement}), and the transient T-MOKE asymmetry now reliably reflects the ultrafast magnetic behaviour.

In summary, we can already conclude that angles of $\sphericalangle(\ptheta{}, \vexy{}) \approx 90 \degree$ can lead to results in the time-resolved data that are not straightforward to interpret. However, this particular situation can readily be identified, if the asymmetry in a certain spectral range is highly sensitive to the EUV angle-of-incidence and flips sign or approaches zero.

\subsection{Beyond the bulk approximation}
In the previous analysis, we have focused specifically on a single vacuum/magnet interface, as the Fresnel equations for such a case are comparatively simple. However, the magnetic samples studied in ultrafast magnetism, including also this study, are often best described as multilayer structures. In particular, a capping layer is often included to protect the magnetic material from the oxidizing environment; and if no capping layer is included, commonly a native oxide layer forms. For such a sample, expressions for the magnetization-dependent reflectivity can be derived using the transfer matrix formalism \cite{zak1990universal, zak1991magneto, schick2014udkm1dsim}. In the present study, we have implemented the transfer matrix formalism from Ref.~\onlinecite{zak1991magneto} using a symbolic math package (SymPy) \cite{sympy_publication} for Python to calculate the reflectivity, the T-MOKE asymmetry, and the probe vectors \ptheta{} without relying on assumptions concerning the strength of the magneto-optic response. 

\section{Determination of dielectric tensor from T-MOKE data}
\label{sec:experimental_permittivity}
The strong angle-dependence of the T-MOKE asymmetry can be leveraged in order to extract the complex-valued off-diagonal dielectric tensor element \exy{} \cite{zusin2018direct, hochst_magnetic_1997, turgut_stoner_2016}. Here, we present a unique approach to such magneto-optical reflectometry where only a small range of incidence angles needs to be measured for complete access to \exy{} at the given photon energies. Specifically, this approach takes advantage of the strong enhancement and dependence on the incidence angle of the T-MOKE asymmetry around 45\textdegree{}, and furthermore makes use of the well-known non-magnetic components of the dielectric tensor, which allows us to reliably calculate the probe vectors. This approach provides two crucial advantages: the experimental setup can be integrated into a typical EUV T-MOKE setup without requiring a major overhaul, and since only a small number of incidence angles needs to be measured, full femtosecond time-resolved magneto-optical spectroscopy is possible with only a small (less than 5-fold) increase in the measurement time. 

In the following, we will first discuss the extraction of the static off-diagonal dielectric tensor element, which is necessary in order to calibrate certain experimental parameters, before we continue with the extraction of the time-resolved dielectric tensor.

\subsection{Extraction of the static off-diagonal dielectric tensor element}
\label{sec:exy}
Several approaches exist to extract the magneto-optical dielectric tensor element \exy{} at extreme ultraviolet wavelengths \cite{willems_magneto-optical_2019, Valencia_2006, hochst_magnetic_1997, Zusin2018, turgut_stoner_2016}. For thin samples that are best probed in reflection, incidence-angle-dependent or polarization-dependent reflectivity measurements have demonstrated their value \cite{hochst_magnetic_1997, Zusin2018, turgut_stoner_2016}. As these techniques rely on reflectivity measurements for linearly polarized light, they are commonly implemented in laboratory-scale experiments based on high-harmonic generation. A limiting factor for the implementation of these techniques, however, is that they typically require a dedicated experimental geometry. Consequently, there is a need for an \exy{} measurement technique that is compatible with, and can be implemented in, currently existing EUV femtomagnetism experiments. 
Here, we present the magneto-optical reflectometry method that we implemented in our table-top femtosecond EUV T-MOKE spectroscopy setup \cite{moller_ultrafast_2021}. This was made possible by the implementation of a motorized, non-magnetic and UHV-compatible tip-tilt sample holder (SmarAct GmbH), which enabled a scan range of 2.5\textdegree{} (see Fig.~S1 in the Supplemental Material (SM)). As such, we anticipate that the method can be rapidly implemented in many setups that are currently being used for fixed-angle EUV T-MOKE. 

\begin{figure}[htb!]
\centering
    \includegraphics[width=0.5\textwidth]{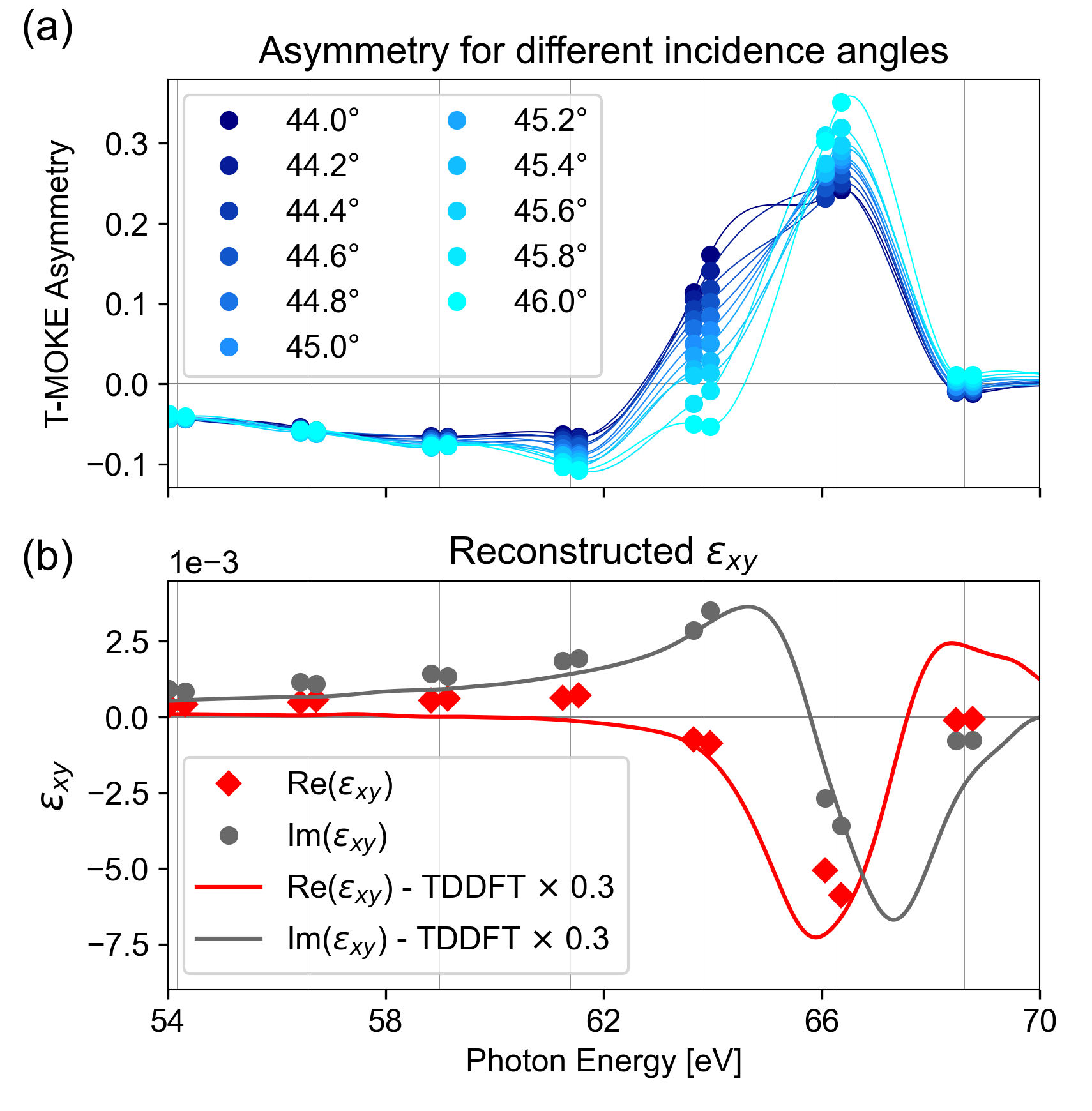}
    \caption{Determination of a static \exy{}. (a) T-MOKE asymmetry for 11 incidence angles close to 45\textdegree{}. Due to the high resolution of the spectrometer, we were able to evaluate the T-MOKE asymmetry at two energies for each high harmonic, indicated by points. The connecting lines serve as a guide to the eye for the T-MOKE asymmetry. Note that only the relative angles were directly determined from the sample, while the absolute angle was retrieved from the analysis. 
    (b) For comparison, we have calculated the real and imaginary parts of \exy{} by TDDFT. We find a good agreement with regard to the photon energy dependence, although TDDFT predicts an overall stronger magneto-optical response. 
    }
    \label{fig:static_exy}
\end{figure}

In order to gain access to the off-axis dielectric tensor element of a magnetized Ni thin film, we performed a series of T-MOKE asymmetry measurements at 11 different incidence angles (shown in Fig.~\ref{fig:static_exy}a). The sample consists of a 21~nm Ni film that was deposited by sputtering on a Si\textsubscript{3}N\textsubscript{4}-coated Si wafer. This thickness was chosen such that femtosecond optical pulses will excite the sample homogeneously, while simultaneously being thick enough that the reflection at the Ni/Si\textsubscript{3}N\textsubscript{4} interface does not modify the observed T-MOKE asymmetry strongly. The 100~nm Si\textsubscript{3}N\textsubscript{4} layer is thick enough that no reflection from the Si\textsubscript{3}N\textsubscript{4}/Si interface has to be accounted for. Finally, the Ni layer possesses a thin, native Ni oxide layer on top, which reduces the strength of the T-MOKE asymmetry. In line with literature \cite{uchikoshi_study_1994, zink_efficient_2016, wang_microstructure_2011}, we set the thickness of the NiO layer to 2~nm. In our analysis, we then use the transfer matrix formalism and calculate expressions for the full three-layer system, including the NiO capping layer, the Ni magnetic layer and the Si\textsubscript{3}N\textsubscript{4} bottom layer.

For our sample, the T-MOKE asymmetry depends on the complex-valued refractive indices of Ni, NiO and  Si\textsubscript{3}N\textsubscript{4}, the layer thicknesses, the incidence angle, and the complex-valued \exy{} of Ni. Using the result of the transfer matrix formalism as the model for the sample, the energy-resolved complex-valued \exy{} can be determined from a set of angle-dependent T-MOKE asymmetry measurements using a standard least-square fitting procedure. In order to facilitate a reliable determination of \exy{} with few measurements over a small angular range, we fix the complex-valued refractive indices to literature values. Here, we retrieve the values for the dispersive (real) part of the refractive index $\delta$ from CXRO \cite{henke_x-ray_1993}, while we used more recent measurements from Ref.~\onlinecite{willems_magneto-optical_2019} for the absorptive (imaginary) part $\beta$. Using other sources for $\beta$ such as Refs.~\onlinecite{Valencia_2006} or \onlinecite{henke_x-ray_1993} leads to small variations in the overall amplitude of \exy{}, but does not affect the photon-energy dependence significantly.

The best-fit values for the real and imaginary part of the off-diagonal tensor element \exy{} in Ni are shown in Fig.~\ref{fig:static_exy}b. As expected, the largest values can be found around the M absorption edges of Ni around 66~eV (M$_3$ edge). We also find a good qualitative agreement with previously published data \cite{Valencia_2006, willems_magneto-optical_2019}. Furthermore, the dielectric tensor can be accurately calculated based on (time-dependent) density functional theory complemented by calculations in the $GW$ framework to achieve an accurate description of the 3p core states \cite{krieger2015laser, dewhurst2020element, willems_magneto-optical_2019, dewhurst_elk}. Here, we find a good agreement between experiment and theory on the shape of the dielectric tensor, although theory indicates an overall larger amplitude of \exy{}. We attribute the discrepancy in the amplitude to a common overestimation of the film-averaged magnetic moment in the TDDFT calculation \cite{jana_analysis_2020, Gang_2018_exy_NiPd}. 
In this regard, a better match between experiment and theory, particularly for the spectral region just above the edge, was recently also achieved by manually reducing the exchange splitting \cite{willems_magneto-optical_2019}.

\subsection{Systematics and uncertainties}
There are several experimental uncertainties that may influence the extracted values for \exy{}. These include statistical and systematical effects. The statistical effects include shot-to-shot intensity fluctuations, position-dependent detection efficiency on the camera and an uncertainty in the relative angle determination. As we performed a large number of angle-dependent measurements with a relatively long measurement time, however, we estimate that the statistical errors are negligible compared to the systematic ones. 
We have identified several possible sources of systematic errors, which we analyse in the context of the present Ni sample:

First, the leading source of systematic errors is the calibration of the incidence angle. From Fig.~\ref{fig:asymm_arrow}, it is clear that a change in the incidence angle leads to a rotation of the probe vectors, as well as a change in the length. Therefore, a systematic error in the angle determination leads to a rotation of \vexy{} in the complex plane, as well as a change in amplitude. Performing the full \exy{} reconstruction for different angle calibrations, we find that a 0.5\textdegree{} calibration error in the incidence angle typically leads to a rotation of \vexy{} by 10\textdegree{} and a scaling of the amplitude by 10\%. Considering the experimental setup, we estimate that the absolute angle calibration is accurate to within 0.3\textdegree{}.

Second, quantitative determination of the T-MOKE asymmetry relies on an accurate subtraction of the background signal in the spectrometer. This is particularly critical when the detected EUV flux is low. To avoid errors due to this effect, we have chosen to evaluate the data only at photon energies around the peaks of the high-harmonic generation (HHG) spectrum. Nevertheless, we note that close to the HHG cut-off energy ($\approx$~70~eV, cf. Fig.~S2 in SI), the flux of each harmonic significantly decreases. At these photon energies, an imperfect background correction can lead to a reduction of the observed T-MOKE asymmetry and thereby an underestimation of the amplitude of \exy{}. For the data reported in this paper, we estimate this effect to be negligible.

Third, it was already discussed that the NiO capping influences the observed T-MOKE asymmetry. Unfortunately, no direct determination of the capping layer thickness was possible. Therefore, we have analysed the influence of the capping layer thickness on the retrieved \exy{} values. Within the range of expected NiO thicknesses (1.5 to 2.5~nm), we find that a thicker (thinner) NiO capping would reduce (increase) the size of the observed T-MOKE asymmetry and therefore would lead to a slightly larger (smaller) amplitude of \exy{}. Here, 0.5~nm change corresponds to less than 10\% change in the amplitude. Also, a comparatively small rotation of the extracted \exy{} can be observed. Overall, the effect due to uncertainty in the capping layer thickness is less pronounced than the effect due to the angle of incidence calibration. 

Fourth, we note that the presented \exy{} reconstruction technique depends on accurate values for the non-magnetic refractive indices of all materials in the sample. However, by performing the full analysis for different values of the refractive index of Ni (from Refs.~\onlinecite{henke_x-ray_1993, Valencia_2006, willems_magneto-optical_2019}), we find only a minor effect on the shape and strength of \exy{}. Related to that, also an accurate calibration of the photon energy is necessary. Within the experimental uncertainty of the photon energy calibration, which is $<2\%$, we find no significant change in the reconstructed \exy{}.

\subsection{Probing the transient dielectric tensor}
\label{sec:transient_exy}
Next, we proceed to determine the time-resolved \exy{} during optically-induced ultrafast demagnetization with sufficient time resolution to trace the full demagnetization process and thereby provide an ideal basis for comparison with theoretical methods that describe the dynamics of the electron, spin and lattice degrees of freedom. 
Compared to the static case that was previously discussed, we make two observations: First, as a consequence of the optical excitation, it is not possible to fix the refractive index to literature values; rather, the dynamical behaviour of $\beta$ and $\delta$ (see Eq.~\eqref{eq:refractive_index}) must be extracted from the experimental data. Second, the number of measured angles must be limited as much as possible to facilitate the measurement of a larger number of pump-probe delays. We find that these points can be addressed by choosing the reflectivity $R_{\uparrow/\downarrow}$ (Eq.~\eqref{eq:intensity_up_down}; more precisely its equivalent from the transfer matrix formalism) as a starting point for the time-resolved analysis, rather than the T-MOKE asymmetry (Eq.~\eqref{eq:asym}). Instead of a single value per photon energy and pump-probe delay, this yields two measurement values (for two magnetization directions) which contain both magnetic and non-magnetic contributions. 

\begin{figure}[tbh]
\centering
\includegraphics[width=\columnwidth]{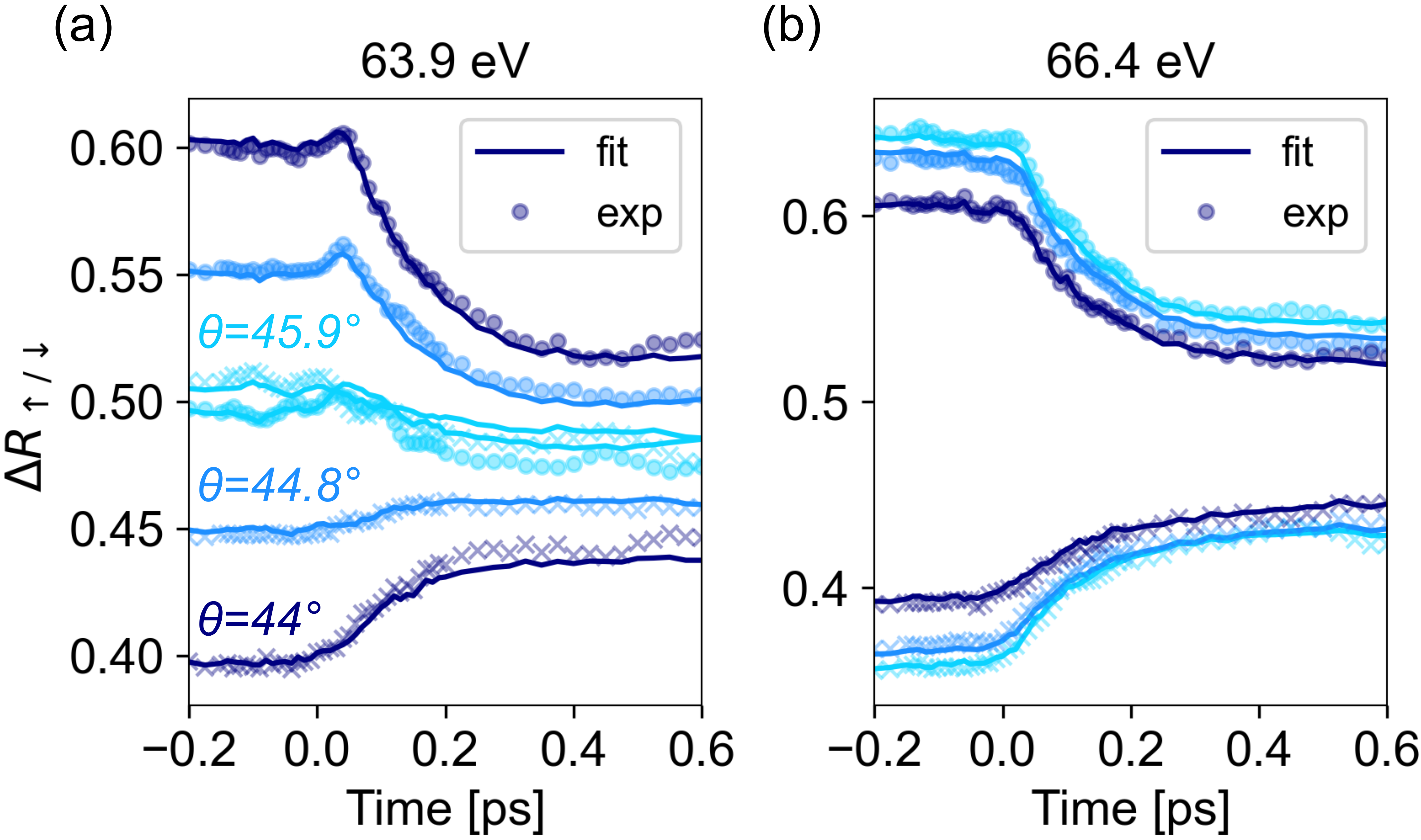}
\caption{
    Fitting of the time-resolved data to extract both \exy{} and $\beta$. (a, b) The measured (points) and reconstructed (line) transient relative reflectivity $\Delta R_{\uparrow/\downarrow}(t)$ for opposite magnetization directions (cf. Eq.~\eqref{eq:rel_reflectivity}) during optically-induced magnetization dynamics for photon energies just below (a) and close to (b) the Ni M-edge.
}
\label{fig:tr_reflectivity}
\end{figure}

\begin{figure}[t!]
\centering
    \includegraphics[width=0.9\columnwidth]{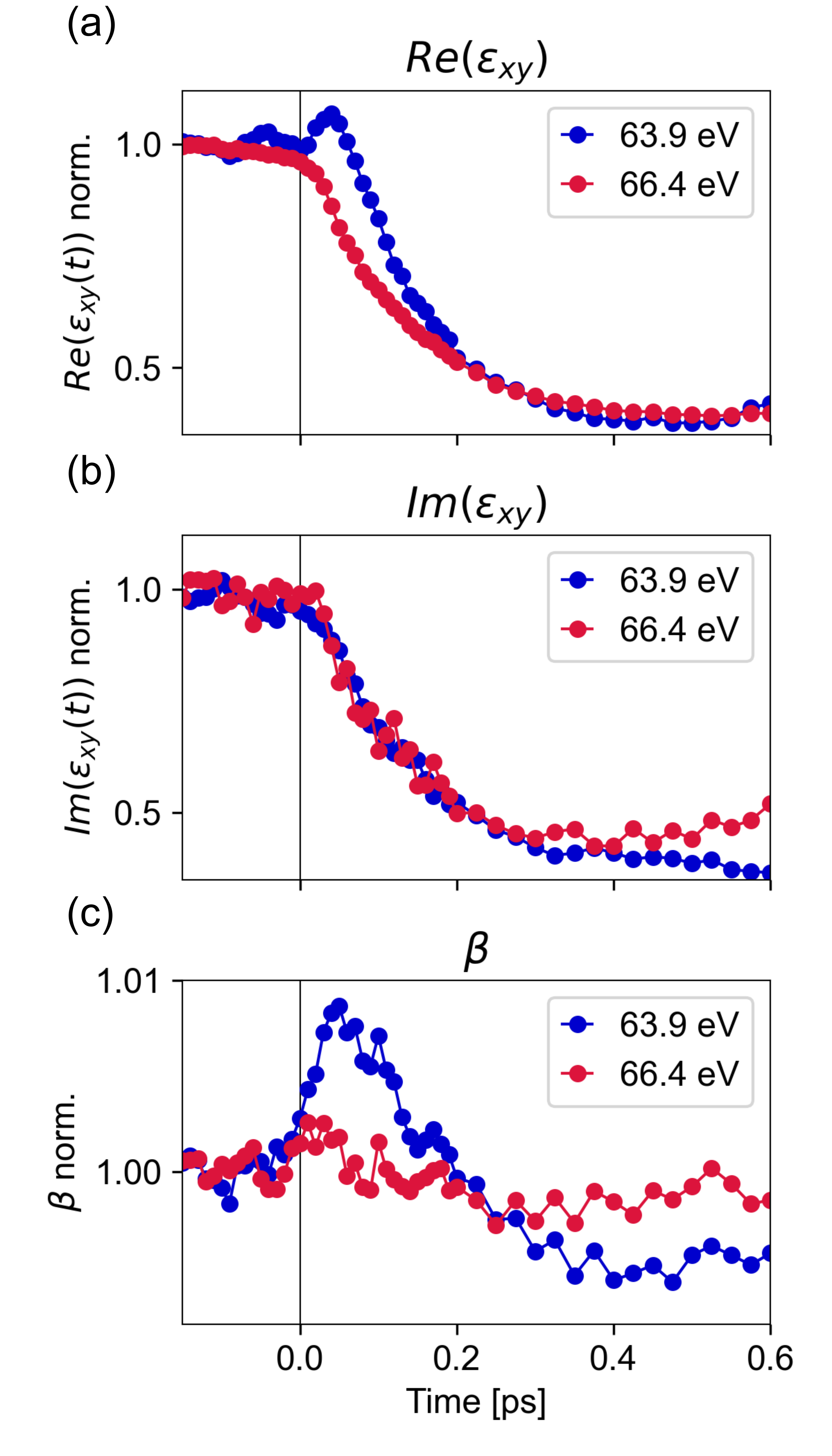}
    \caption{(a, b) Transient evolution of \exy{} after the optical excitation. \exy{} rotates in the complex plane, because the real and imaginary part do not change identically. The rotation of \exy{} is of particular importance to understand the differences in the asymmetry time-traces for different incidence angles below the resonance. (c) Transient changes of the imaginary part of the refractive index ($\beta$) of Ni retrieved from the angle-resolved EUV T-MOKE data. We find a strong change of $\beta$ at $h\nu =63.9$~eV that is indicative for the strong excitation of electrons. }
    \label{fig:exy_rotation}
\end{figure}

\begin{figure*}[ht!]
\centering
    \includegraphics[width=0.9\textwidth]{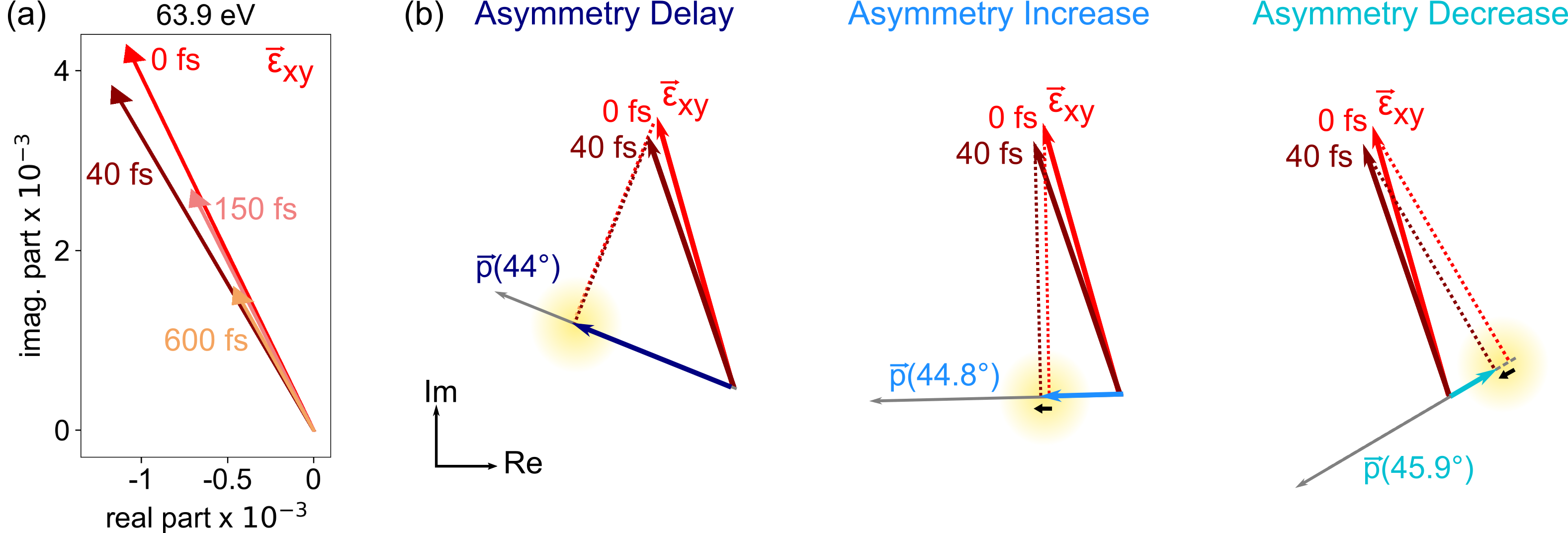}
    \caption{(a) Transient evolution of \vexy{} for selected times after the optical excitation. \vexy{} rotates in the complex plane, as the real and imaginary part do not change identically. 
    (b) A rotation of \vexy{} in the complex plane can lead to an increase, a decrease, or no effect on the magnetic asymmetry, which can be understood from the alignment of \vexy{} and \ptheta{}. Here, we specifically consider the change in \vexy{} of Ni at 64~eV and 40~fs after the onset of demagnetization. At 44\textdegree{}, the rotation of \vexy{} is such that the projection on \ptheta{} does not change. Hence, no change in the magnetic asymmetry is observed. At 44.8\textdegree{}, the rotation leads to an increase of the projection (magnetic asymmetry), while at 45.9\textdegree{}, a pronounced decrease is observed. The change of the projection from \vexy{} on \ptheta{} between unpumped and at 40~fs is highlighted in yellow.}
    \label{fig:exy_rotation2}
\end{figure*}

As the absolute reflectivity cannot be measured in our experiment, we consider only the relative changes compared to the unpumped ($t<0$) case. From a measurement series of the delay-dependent intensity $I_{\uparrow/\downarrow}(t)$, we extract the relative reflectivity
\begin{equation} \label{eq:rel_reflectivity}
    \Delta R_{\uparrow/\downarrow}(t) = \frac{1}{2} \frac{R_{\uparrow/\downarrow}(t)}{R_{\uparrow/\downarrow}(t_{\rm ref})} = \frac{I_{\uparrow/\downarrow}(t)}{I_{\uparrow}(t_{\rm ref}) + I_{\downarrow}(t_{\rm ref})},
\end{equation}
where $t_{\rm ref}$ indicates the reference time interval where the probe arrives at the sample before the pump pulse. This yields a signal that reflects the static asymmetry at times before the pump pulse and provides a measure of the transient magneto-optical and non-magnetic reflectivity changes at times after the pump pulse has arrived. Fig.~\ref{fig:tr_reflectivity}a and b show such measurement data for the same experimental situation that was presented in Fig.~\ref{fig:dynamics_angles}, namely the optically-induced demagnetization of Ni. 

The full data set contains the magnetization-dependent relative reflectivity for the full high-harmonic spectrum at 3 incidence angles $\theta_i$ (44\textdegree{}, 44.8\textdegree{} and 45.9\textdegree{}, as determined from the static analysis). In order to extract the time-dependent \exy{}, we apply a similar fitting procedure as was used in the static case: Each photon energy and each pump-probe delay are considered independently. For each combination of these, our reconstruction algorithm determines the optimal values for \reexy{}, \imexy{}, $\beta$ and $\delta$. 

We find that $\delta$ generally does not influence the predicted reflectivity strongly, and consequently it cannot accurately be determined with a least-square fitting routine. To address this, we employ a self-consistent two-step approach: 1. keeping $\delta$ fixed, we use a least-square fitting to determine \reexy{}, \imexy{} and $\beta$. 2. We use a Kramers-Kronig (KK) transform to determine $\delta$ \cite{watts_calculation_2014}. Repeating steps 1 and 2 up to ten times has shown excellent convergence and a good reconstruction of the measured values, as shown in Fig.~\ref{fig:tr_reflectivity}a and b. We note that although \reexy{} and \imexy{} are also connected by the KK relations, the implementation of such a transform is challenging due to the sparsely sampled nature of our spectrum. We have therefore achieved better results without linking \reexy{} and \imexy{} by KK.


\section{Interpretation}
\subsection{Rotation of \exy{}}

\label{sec:rotation_exy}
Having reconstructed the transient dielectric tensor including both the magnetic (\exy{}) and non-magnetic (\exx{}) part, we proceed with the interpretation of these results. In a first step, it is instructive to return to the special case that was identified in Ni at 64~eV photon energy (Fig.~\ref{fig:dynamics_angles}). Fig.~\ref{fig:exy_rotation}a and b show the real and imaginary part of the off-diagonal dielectric tensor element, respectively, as a function of time for two selected energies (more data, including the extracted variation of $\delta$, are shown in Fig.~S5 in the SM). First, we recognize that the overall magnitude $|\exy{}|$ decreases over time to approximately 40\% of its original length after 600~fs. More interesting, however, we also see that \reexy{} and \imexy{} do not quench identically as a function of time, especially for $h\nu=$~64~eV, where a transient increase of \reexy{} is observed. This means that \exy{} must transiently rotate in the complex plane. This rotation in the complex plane is visualized in Fig.~\ref{fig:exy_rotation2}a and is a key factor in the explanation of the diverse transient effects that were observed in Ni at 64~eV (Fig.~\ref{fig:dynamics_angles}). However, we note that a rotation of \exy{} can also be observed for many other photon energies, and does not necessarily have a large impact on the T-MOKE asymmetry data, as is evident from the measurement shown in Fig.~\ref{fig:good_measurement}. Therefore, a rotation of \exy{} alone is not sufficient to explain the peculiar data of Fig.~\ref{fig:dynamics_angles}.

At 64~eV, we find that, in addition to the transient rotation of \exy{}, the probe vector \ptheta{} and the off-diagonal dielectric tensor element \exy{} are close to orthogonal. This is evident by the zero-crossing of the static magnetic asymmetry for different angles-of-incidence (see Figs.~\ref{fig:asymm_arrow} and \ref{fig:static_exy}). This situation does not occur at the other photon energies that we have observed, and it provides the second key to the explanation of the data in Fig.~\ref{fig:dynamics_angles}. The T-MOKE asymmetry can be approximated by the projection of \exy{} on the probe vectors \ptheta{}. In Fig.~\ref{fig:exy_rotation2}b, we now apply this method again to visualize how the experimentally-determined transient rotation of \exy{} can influence the measured T-MOKE asymmetry. We find that changes of the orientation of \exy{} can lead to dramatic effects, which furthermore depend strongly on the precise angle $\sphericalangle(\vec{p}, \vexy{})$, and in particular if this angle is more or less than 90$\degree$. We can readily construct scenarios that predict a delay, a rapid increase and a rapid decrease in the T-MOKE asymmetry - exactly as was observed for Ni (cf. yellow areas in Fig.~\ref{fig:exy_rotation2}b). Furthermore, we note that a complete reversal of the T-MOKE asymmetry might also occur. A naive interpretation of the T-MOKE asymmetry might confuse this observation for a reversal of the magnetic moment, which we emphasize is not the case here. 
Thus, we conclude that in experiment, where the transient change of the T-MOKE asymmetry is always normalized to its value before the pump excitation, even very small angle changes of \exy{} can have a huge impact on the qualitative trend of the transient T-MOKE asymmetry data.

Since the T-MOKE asymmetry depends on the relative angle $\sphericalangle(\vec{p}, \vexy{})$, it is clear that not just a rotation of \exy{}, but also a rotation of \ptheta{} might lead to unexpected dynamics in the time-resolved T-MOKE asymmetry signal. Since such rotations can arise due to changes of the refractive index, they are commonly referred to as non-magnetic artifacts in the ultrafast magnetism community \cite{Koopmans_2003, la-o-vorakiat_ultrafast_2012}. However, we note that in our time-resolved analysis, $\beta$ (Fig.~\ref{fig:exy_rotation}c) and $\delta$ (Fig.~S5 in SM) are reconstructed in addition. Therefore, this analysis allows to separate the rotation of \exy{} from any rotation in \ptheta{}. Equivalently, this analysis separates the magnetic and non-magnetic effects. For Ni at 64~eV, we find that the \ptheta{} vectors change insignificantly in comparison to \exy{}, indicating that the angle-dependent T-MOKE asymmetry changes are dominated by the rotation of \exy{}.

In summary, we find that the optically-induced non-equilibrium excitation in Ni leads to a transient rotation of the off-diagonal element of the dielectric tensor. In consequence, we show that the T-MOKE asymmetry may exhibit increased, decreased, and delayed behavior, which must not be directly interpreted as transient dynamics of the magnetization of the sample.

\subsection{Comparison to theory}
\label{sec:rotation_exy_theory}
\begin{figure*}[htb!]
\centering
    \includegraphics[width=\textwidth]{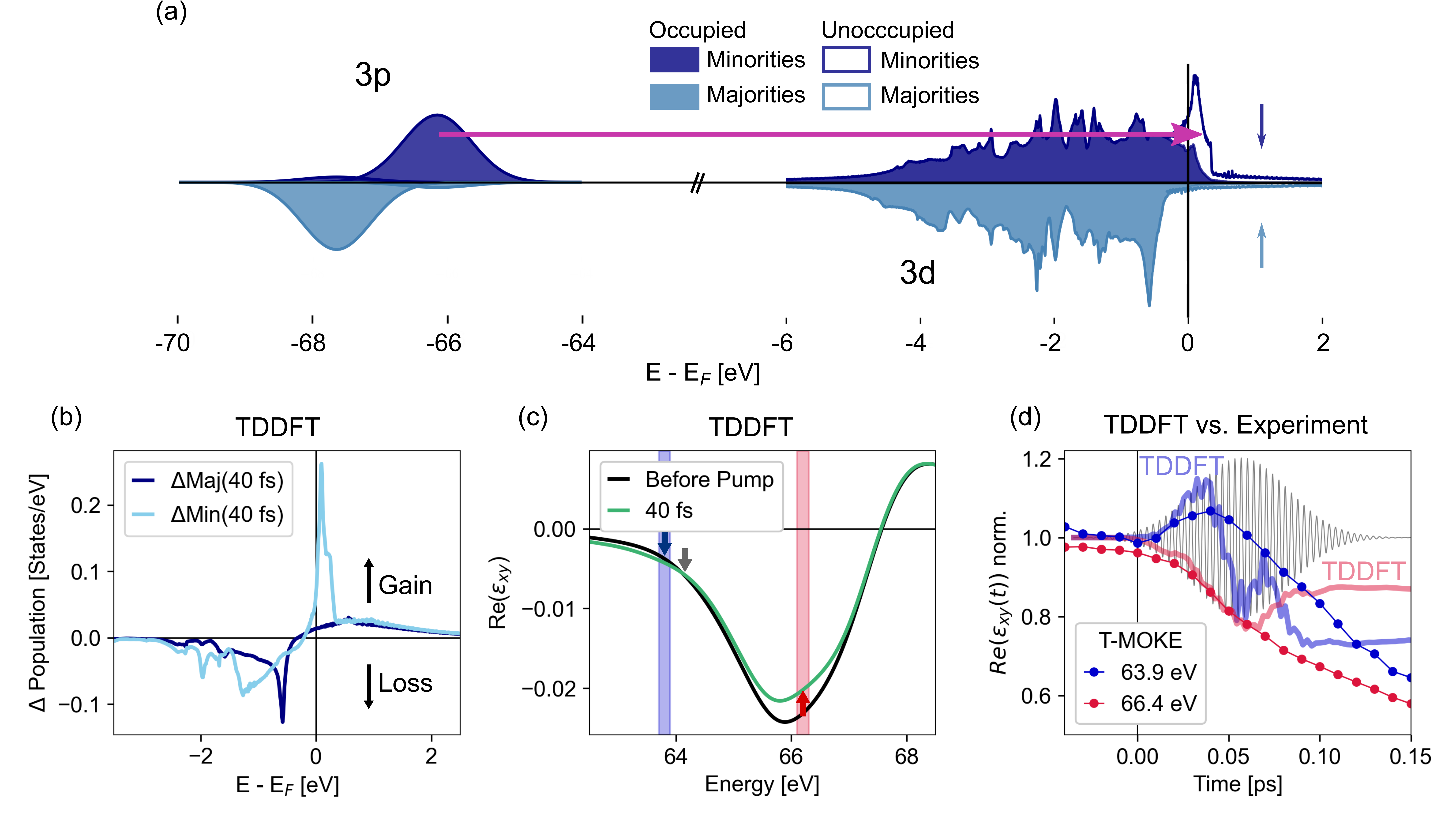}
    \caption{Femtosecond spin dynamics from TDDFT, compared to experiment. (a) Spin-resolved density of states from density functional theory and the equilibrium occupation at 300~K for the 3p, 3d and 4s states in Ni. Due to the exchange splitting, significantly more minority states (upper part) are unoccupied (unshaded region around $E_F$), thereby allowing for more optical transitions (purple arrow).
    (b) As seen in the population change at 40~fs, the optical excitation with 1.2~eV photons leads to a strong redistribution of the population in the minority channel (light blue), populating empty states at 0-0.5~eV above $E_F$ and creating empty states below $E_F$. In comparison, the majority channel (dark blue) is less affected. 
    (c) The redistribution of spins and charge carriers leads to a modification of the magneto-optical response (\exy{}). Depending on the position with respect to the M-edge photon energy $E_M$, the spectral dynamics of \reexy{} show regions with a strong relative decrease of \reexy{} (indicated with a red arrow) but also regions with a relative increase of \reexy{} (blue arrow) or a delayed onset of dynamics (grey arrow).
    (d) The calculations can be directly compared to experiment by evaluating \reexy{} as a function of time. Here, we observe that the predicted relative increase in \reexy{} is confirmed from experiment (see Fig.~\ref{fig:tr_reflectivity}a). The time axis of experiment and theory were shifted according to each, as visualized by the electric field of the pulse used for the theory calculations. 
    }
    \label{fig:Real_exy_exp_theory}
\end{figure*}

Before we carry out a direct comparison to theory, it is instructive to have a qualitative look at the resonant M-edge transitions that are probed in experiment with EUV T-MOKE spectroscopy. Fig.~\ref{fig:Real_exy_exp_theory}a shows a schematic of the spin-split 3d and 4s density of states of Ni and the spin-split Ni 3p core levels, for which we include the approximate intrinsic linewidth broadening. As can be seen from this schematic, the spin-split 3p core levels overlap and cover an energy range of $\approx$5~eV, which is comparable to the energy range of the full valence and conduction band structure of Ni. Therefore, a specific EUV energy (pink arrow in Fig.~\ref{fig:Real_exy_exp_theory}a) does not only probe a specific energy within the spin-split DOS, but an extended region of several electronvolts. In consequence, the collected T-MOKE asymmetry is strongly broadened, and it is not straightforward to identify spectrally-resolved dynamics in the T-MOKE asymmetry or even the extracted dielectric tensor with occupation changes or band renormalizations in the electronic structure of the investigated material.

However, if we look at our TDDFT calculations in Fig.~\ref{fig:Real_exy_exp_theory}b, such a comparison seems necessary at first. For 1.2~eV and 47~fs pump pulses, we find that the time-resolved change in minority and majority occupation of the DOS exhibits spectrally very distinct dynamics. For example, because of the large unoccupied DOS in the minority channel of Ni just above the Fermi-level (cf. Fig.~\ref{fig:Real_exy_exp_theory}a), Fig.~\ref{fig:Real_exy_exp_theory}b shows that pumping with 1.2\,eV leads to a peaked increase of minority spins in a $\approx$300~meV spectral region above the Fermi level. In other words, we see a strong minority inter-energy transfer of spins from below to above the Fermi-level. In the majority channel, we also pump electrons from below to above the Fermi-level, but much less efficient, because there are much less empty states available for the transition. In total, this means that the optical excitation leads to a magnetic moment increase below the Fermi-level (loss of minority electrons), and an overall magnetic moment decrease (gain of minority electrons) just above the Fermi-level. However, all of these distinct spectral signatures will be broadened in experiment due to the large linewidth of the Ni 3p core levels. 

We can overcome this problem by directly calculating the transient changes in the dielectric tensor, which takes all EUV transitions into account and moreover is a quantity that we now can directly compare with experiment. Specifically, we focus on the real part of \exy{}, which can be related to the spin polarization of the unoccupied states. Fig.~\ref{fig:Real_exy_exp_theory}c shows the theoretical transient dynamics of \reexy{} in the corresponding energy range. First, we recognize that the spectrally distinct dynamics as seen in the transient occupation changes in Fig.~\ref{fig:Real_exy_exp_theory}b are completely smeared out. Nevertheless, we still find broad energetic regions where the inter-energy spin transfer as seen in Fig.~\ref{fig:Real_exy_exp_theory}c can be verified: Below the energy marked with the grey arrow in Fig.~\ref{fig:Real_exy_exp_theory}b, \reexy{} shows an ultrafast relative increase, and an ultrafast decrease for energies above. If \reexy{} would be extracted exactly at the energy of the green arrow, we would expect a delayed behavior in the dynamics. 

Having this theoretical result on transient dynamics of \reexy{}, we now carry out a direct comparison with experiment, where we were able to extract \reexy{} at the energies marked with blue- and red-shaded areas in Fig.~\ref{fig:Real_exy_exp_theory}c. Figure~\ref{fig:Real_exy_exp_theory}d shows this comparison of time-resolved dynamics of \reexy{} for theory (solid lines) and experiment (data points). Clearly, the predicted relative increase and decrease in \reexy{} after optical pumping can be confirmed. We also find that for longer time-delays >100~fs, the experimental data shows a further reduction of \reexy{}, while the theory data stays constant. This is to be expected, because TDDFT does not take all possible spin-flip scattering processes into account, which in experiment lead to the overall demagnetization of the sample on timescales larger than 100~fs.

In summary, TDDFT predicts an OISTR-like process, i.e. an optically-driven inter-energy spin transfer similar to the inter-site spin transfer in the usual OISTR process. This pumping of spins happens predominantly in the minority channel and leads to distinct spectrally-dependent magnetization increase and decrease below and above the Fermi-level, respectively. Via a direct calculation of the real part of the off-diagonal element of the dielectric tensor, \reexy{}, we are able to pinpoint the inter-energy transfer in the transient dynamics of \reexy{}. As \reexy{} can be directly extracted from our experimental data, we are able to unambiguously verify the optically-driven inter-energy spin-transfer process in Ni.

\section{Conclusion}
In conclusion, we  have discussed an intriguing effect in EUV T-MOKE magnetism studies, where different transient dynamics were observed under almost identical experimental conditions. This observation conclusively illustrates that it is not always possible to directly compare time-resolved T-MOKE data with time-resolved calculations of the spin-resolved density of states. Here, we show that a quantitative comparison with TDDFT can be achieved by extracting the transient dynamics of the off-axis dielectric tensor element \exy{}. 

We have presented a robust technique to retrieve \exy{} from experimental T-MOKE data with full femtosecond time resolution at little experimental cost. The data that we have achieved solves the controversy that slightly different incidence angles lead to dramatically different dynamics in the T-MOKE asymmetry. Even more, our technique allows to directly reconstruct the magnetic and non-magnetic parts of the refractive index. Studying the prototypical case of femtosecond demagnetization in Ni, we show that this data is ideal for a quantitative analysis and comparison with TDDFT calculations and allows to trace the optically-induced spin and charge dynamics in exceptional detail. We want to emphasize that besides a comparison of the same
quantity, i.e. \exy{}, this approach also ensures that spectral broadening, multiple edges, and overlapping edges from multiple elements in multi-component materials is properly taken into account. 

Beyond this exemplary case, we expect that femtosecond \exy{} spectroscopy will be especially valuable to shed light on the recently discovered OISTR effect and similar femtosecond optically-induced spin dynamics. In Ref.~\onlinecite{moller_verification_2023}, we perform such a study, and discuss the implications for OISTR, which was, after all, first experimentally evidenced by EUV T-MOKE.


\section{ACKNOWLEDGEMENTS}
This work was funded by the Deutsche Forschungsgemeinschaft (DFG, German Research Foundation) - project IDs 399572199 and 432680300/SFB 1456. G.S.M.J. acknowledges financial support by the Alexander von Humboldt Foundation. S.S. and J.K.D. would like to thank the DFG for funding through project-ID 328545488 TRR227 (project A04).

\bibliography{2023_dielectric_tensor}
\section{Supplemental Material}
\label{section:SM}
\renewcommand\thefigure{S\arabic{figure}} 
\setcounter{figure}{0}
\section{Experimental setup and analysis details}
The experimental setup used for the determination of the time-resolved dielectric tensor is based on the EUV T-MOKE setup that was described in Ref.~\cite{moller_ultrafast_2021}. We pumped the samples with a $47\pm5$~fs pulse (Gauss FWHM) with a photon energy of 1.2~eV. The absorbed fluence for the measurements was $0.8\pm0.2$~$\nicefrac{\text{mJ}}{\text{cm}^2}$. The reflected 100 kHz EUV probe beam spans energies between $30-72$~eV.  
\begin{figure}[htb!]
\centering
    \includegraphics[width=0.8\columnwidth]{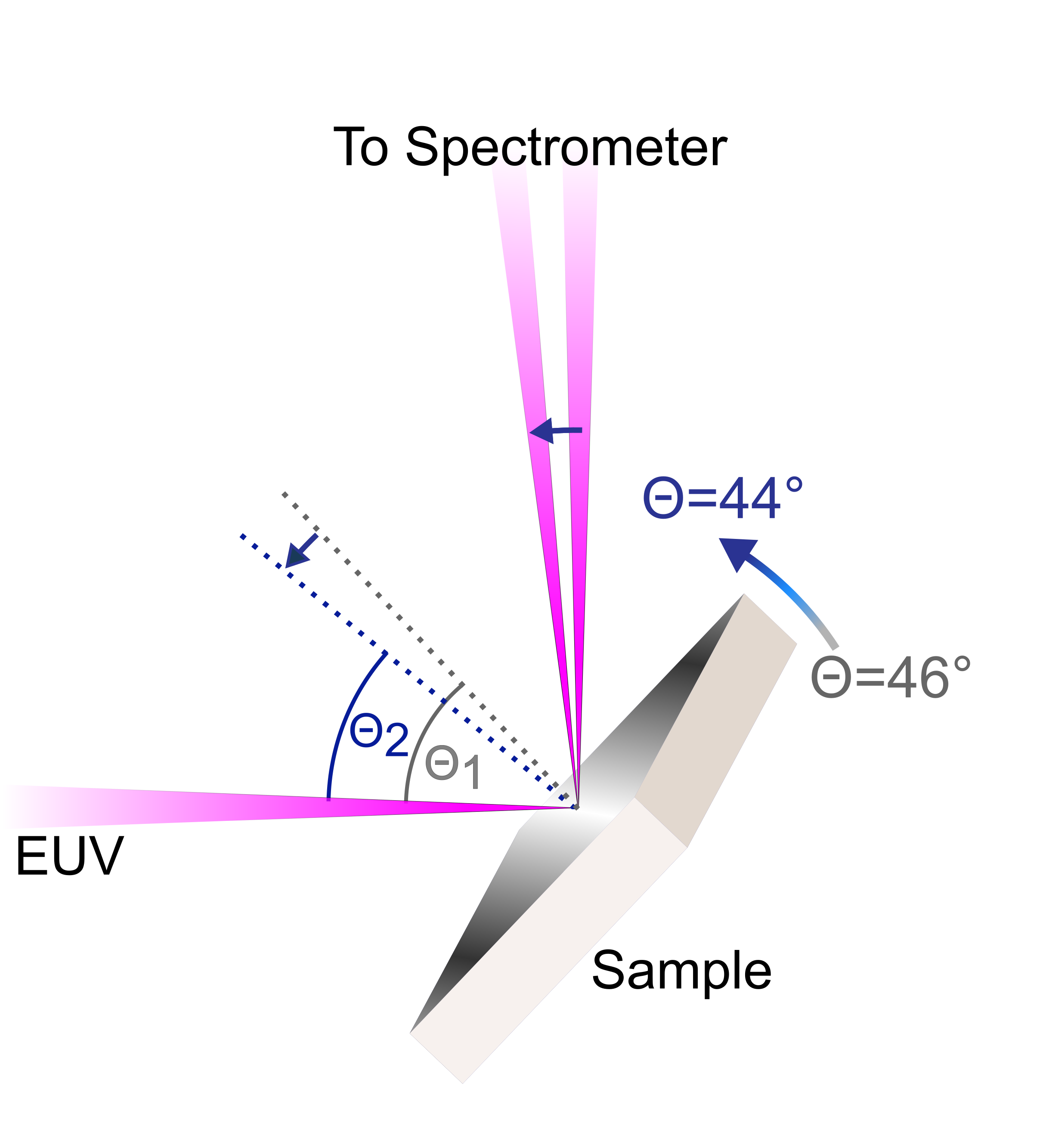}
    \caption{Schematic of the T-MOKE experiment. A small, non-magnetic tip-tilt holder enables repeatable control of the incidence angle in a range of 2.5\textdegree{} around 45\textdegree.   }
    \label{fig:sample_turning}
\end{figure}
For the data acquisition at different angles of incidence (see Fig.~\ref{fig:sample_turning}), an accurate and repeatable control of the incidence angle was ensured by mounting the sample on a small non-magnetic, UHV-compatible tip-tilt stage (SmarAct GmbH, STT-12.7-UHVT-NM, Fig.~\ref{fig:sample_turning}). 

\begin{figure}[htb!]
\centering
    \includegraphics[width=\columnwidth]{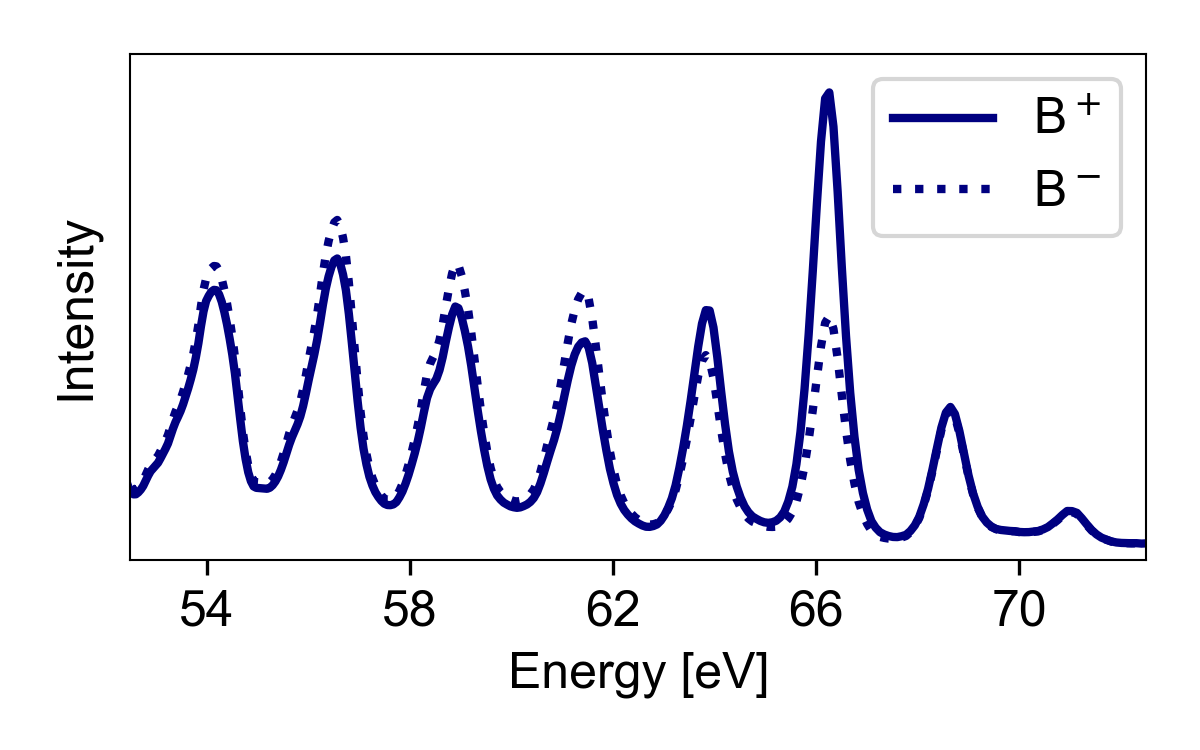}
    \caption{ Typical high harmonic spectrum after the reflection from the Ni thin film for opposite magnetization directions. The discrete nature of the spectrum of odd harmonics of the fundamental beam is clearly visible. The magneto optical response is largest around the Ni resonance at $\approx$66~eV, which is visible by the large difference in intensity for opposite magnetization directions.
    }
    \label{fig:spectrum}
\end{figure}

After reflection from the sample, the EUV light is analyzed in the EUV spectrometer. A typical spectrum is shown in Fig.~\ref{fig:spectrum}. The energy axis of our spectrometer is determined from the spacing of the high harmonics, which is given by twice the fundamental photon energy of 1.2~eV. This leads to a relative uncertainty of $<2\%$ on the photon energy calibration. From the energy-resolved dynamics, we estimate the spectral resolution to be better than 0.2~eV. 

\begin{figure}[htb!]
\centering
    \includegraphics[width=1\columnwidth]{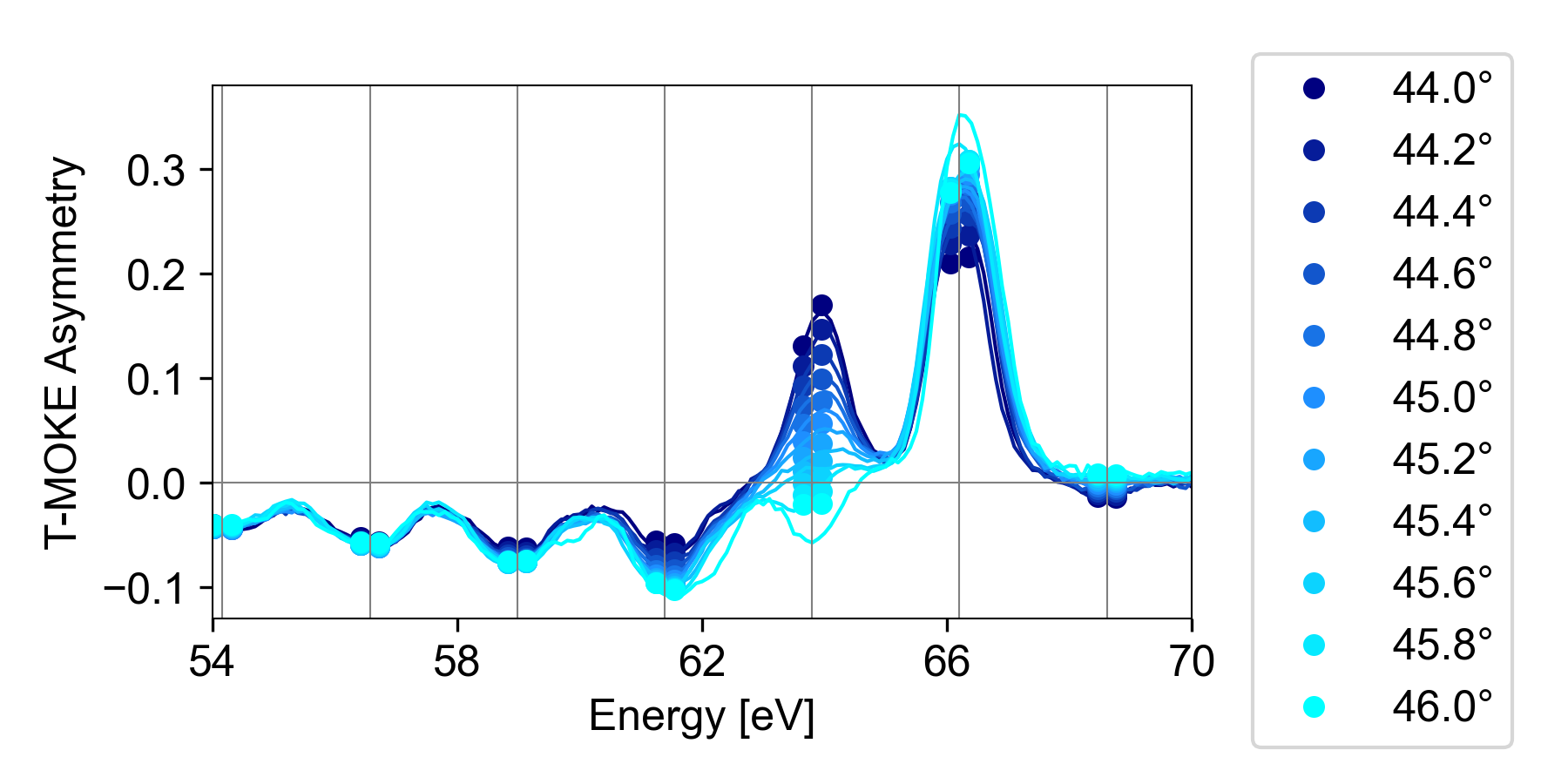}
    \caption{ Fit of the static magnetic asymmetries for the extraction of \exy{} as described in Section III A. The results from the fit procedure (colored circles) provide a satisfactory description of the experimental data (solid lines).  }
    \label{fig:asym_vs_fit}
\end{figure}

For the determination of the off-diagonal dielectric tensor \exy{} from T-MOKE data, the asymmetry was evaluated in 0.1~eV energy windows separated by 0.3~eV (center-to-center) at energies close to the high harmonic maxima. Fit results of the static magnetic asymmetries for the extraction of \exy{} (Section III A) together with the observed T-MOKE asymmetry are shown in Fig.~\ref{fig:asym_vs_fit} and yield good agreement with the measured static T-MOKE asymmetries. We note that the T-MOKE asymmetry is strongly reduced in-between high harmonic energies due to the comparably higher background contribution at those energies. 

\begin{figure}[tbh!]
\centering
\includegraphics[width=\columnwidth]{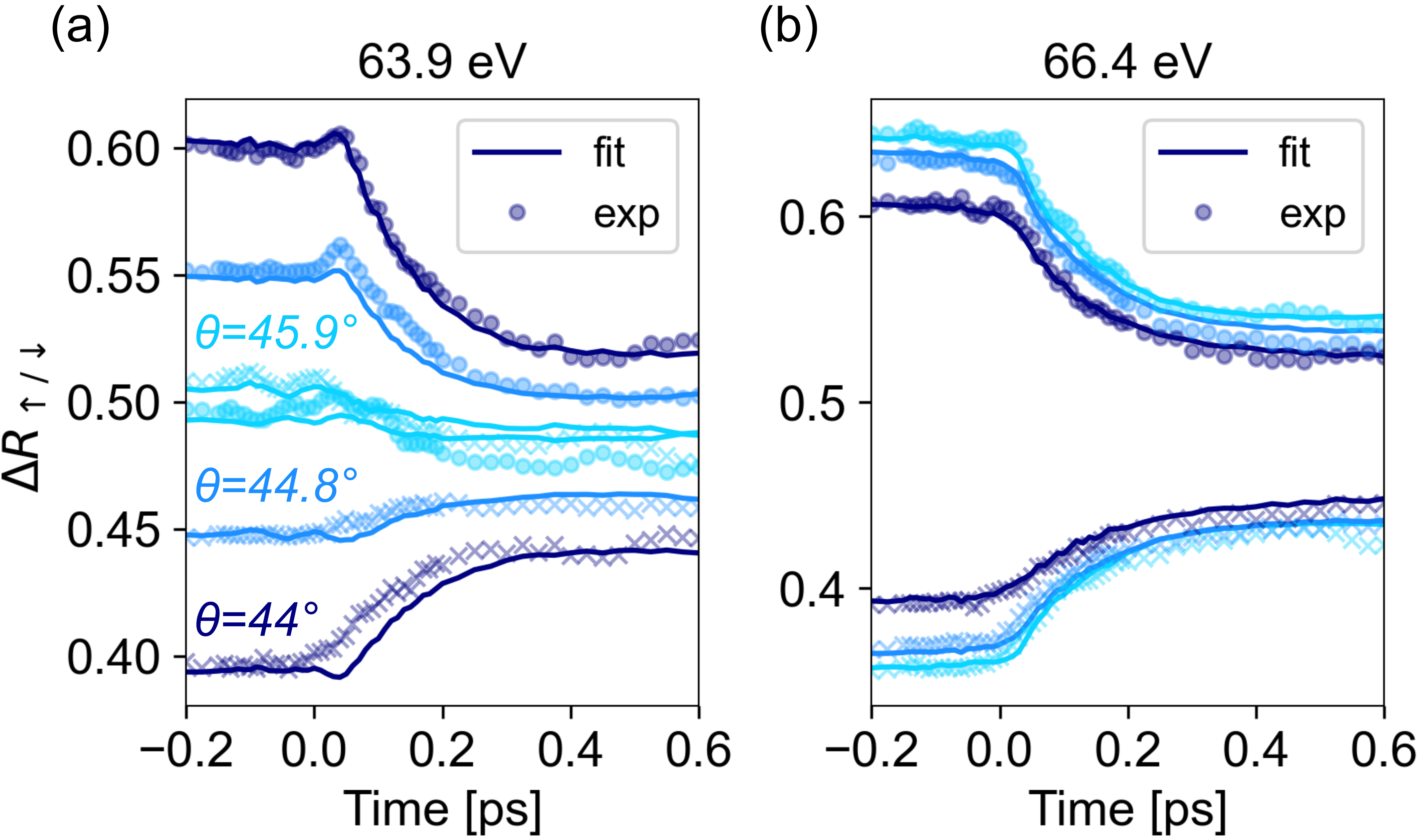}
    \caption{Time-resolved analysis of the dielectric tensor, if transient changes of the refractive index in the sample are neglected. The panels show the measured (points) and reconstructed (line) transient relative reflectivity $\Delta R_{\uparrow/\downarrow}(t)$ for opposite magnetization directions (cf. Eq.~8 in the main text) during optically-induced demagnetization for photon energies just below (a) and close to (b) the Ni M-edge - equivalent to Fig.~6a and b in the main text. The exclusion of transient optical changes of the refractive index results in a poor fit of the experimental data, especially at the energy below the resonance, where the $\Delta R_{\uparrow/\downarrow}(t)$ curves deviate from the usual mirror-symmetric behavior. At this energy, the excitation of electrons by the laser pulse is most dominant and has a strong effect on the refractive index.}
    \label{fig:no_optical_changes}
\end{figure}

\begin{figure*}[htb!]
\centering
    \includegraphics[width=\textwidth]{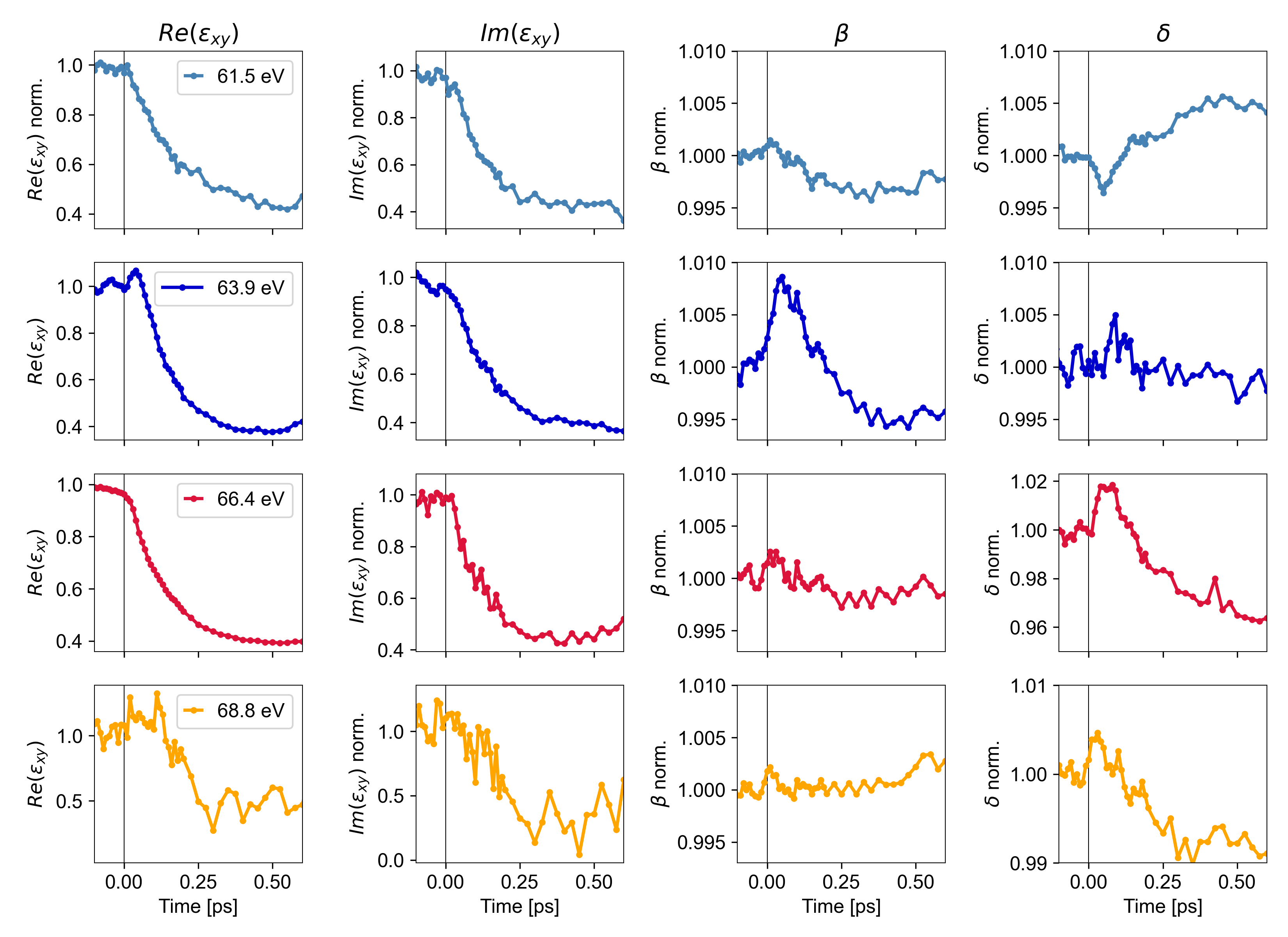}
    \caption{Fit results of the transient behavior of \exy{} and $n$ for different energies around the Ni M-edge.
    }
    \label{fig:fit_results_more_energies}
\end{figure*}

The transient evolution of the off-diagonal dielectric tensor \exy{} was determined from the measured reflectivities $R_{\uparrow/\downarrow}$ (Section III B). For the fitting of $R_{\uparrow/\downarrow}$, good agreement with the experimental data on early timescales can only be achieved by the inclusion of transient optical changes of the complex refractive index $n=1-\delta+i \beta$. In particular, this is important below the resonance (around $\approx64$~eV), where we probe states below \ef{}, which are most strongly affected by the excitation of electrons by the pump pulse. Not including the refractive index changes yields a poor agreement between the measured $R_{\uparrow/\downarrow}$ and the fit, as shown in Fig.~\ref{fig:no_optical_changes}.   
The extracted transient evolution of the real and imaginary part of the off-diagonal dielectric tensor element \exy{} and the optical changes of $\delta$ and $\beta$ for selected energies around the Ni resonance are shown in Fig.~\ref{fig:fit_results_more_energies}.

\section{Time-dependent Density Functional Theory}
For the TDDFT calculations a pump pulse with a central photon energy of 1.2~eV, a Gaussian intensity profile with FWHM = 47~fs and an incident fluence of $\approx$~18~$\nicefrac{\text{mJ}}{\text{cm}^2}$ was used. For the details of the method and the code see Refs.~\cite{krieger2015laser, dewhurst2016efficient, dewhurst_elk}.

\vfill

\end{document}